\documentclass[preprint,showpacs,preprintnumbers,amsmath,amssymb]{revtex4}

\usepackage{graphicx}
\usepackage{dcolumn}
\usepackage{bm}
\usepackage{multirow}
\usepackage{color}

\begin{document}

\title{M$\chi$D candidate nucleus $^{105}$Rh in relativistic
mean-field approach}

\author{Jian~Li}
 \affiliation{School of Physics, State Key Laboratory of Nuclear Physics and Technology, Peking University, Beijing 100871, China}

\author{S. Q. Zhang}
 \affiliation{School of Physics, State Key Laboratory of Nuclear Physics and Technology, Peking University, Beijing 100871, China}

\author{J. Meng}
 \email{mengj@pku.edu.cn}
 \affiliation{School of Physics and Nuclear Energy Engineering, Beihang University, Beijing 100191, China}
 \affiliation{School of Physics, State Key Laboratory of Nuclear Physics and Technology, Peking University, Beijing 100871, China}
  \affiliation{Department of Physics, University of Stellenbosch, Stellenbosch, South Africa}

\date{\today}

\begin{abstract}
Following the reports of two pairs of chiral doublet bands observed
in $^{105}$Rh, the adiabatic and configuration-fixed constrained
triaxial relativistic
mean-field (RMF) calculations are performed to investigate their
triaxial deformations with the corresponding configuration and the
possible multiple chiral doublet (M$\chi$D) phenomenon. The existence
of M$\chi$D phenomenon in $^{105}$Rh is highly expected.
\end{abstract}

\maketitle

Chirality is a topic of general interest in nature science, such as
chemistry, biology and physics. The occurrence of chirality in
atomic nuclear structure was suggested for triaxially deformed
nuclei in 1997~\cite{Frauendorf1997} and the predicted patterns of
spectra exhibiting chirality --- chiral doublet bands --- were
experimentally observed in 2001~\cite{Starosta2001}. In addition to
the triaxial deformation, the configuration with high-$j$ valence
particle(s) and valence hole(s) is also essential for chirality in
nuclei. Up to now, candidate chiral doublet bands have been proposed
in a number of odd-odd, odd-A or even-even nuclei in the $A \sim
100, 130, 190$ mass regions, for a review, see
e.g.~\cite{Meng2010,Meng2010a}.

Theory wise, chiral doublet bands were first investigated in the
one-particle-one-hole-rotor model (PRM) and the corresponding
tilted axis cranking (TAC) approximation~\cite{Frauendorf1997}.
Later on, numerous efforts have been devoted to the development
of TAC~\cite{Dimitrov2000,Olbratowski2004,Olbratowski2006}
and PRM models~\cite{Peng2003,Koike2004,Wang2007,Zhang2007,Qi2009} to describe chiral rotation
in atomic nuclei.

RMF theory has received wide attention due
to its massive success in describing nuclear global properties and
exotic phenomena~\cite{Ring1996,Vretenar2005,Meng2006a}. In order to
describe the nuclear rotation phenomena, the cranked RMF theory in
the context of principal axis
rotation~\cite{Koepf1989,Afanasjev2000} as well as three-dimensional
rotation~\cite{Madokoro2000} were developed. However due to
numerical complexity, it was restricted  to  two-dimensional studies
only, i.e., the magnetic rotation~\cite{Madokoro2000,Peng2008}.

Based on the adiabatic and configuration-fixed constrained triaxial
RMF calculation, triaxial shape coexistence with high-$j$ proton
hole- and neutron particle-configurations, are found in $^{106}$Rh,
which suggests the existence of a new phenomenon --- M$\chi$D~\cite{Meng2006}. This prediction holds true for
other rhodium isotopes $^{104,106,108,110}$Rh as
well~\cite{Peng2008a}. In particular, the prediction of the M$\chi$D
in $^{106}$Rh remains even with the time-odd fields
inclued~\cite{Yao2009}.

Recently, two pairs of chiral doublet bands have been respectively
observed in $^{105}$Rh with three-quasiparticle configurations $\pi
g_{9/2}\otimes \nu h_{11/2}(g_{7/2},d_{5/2})$~\cite{Alcantara2004} and
$\pi g_{9/2}\otimes \nu h^2_{11/2}$~\cite{Timar2004}. It is
interesting to verify whether these observations are the predicted
M$\chi$D or not in the configuration-fixed constrained triaxial RMF
approach. Similar as in Ref.~\cite{Yao2009}, the time-odd fields are
included.


The starting point of the RMF theory is the standard effective
Lagrangian density constructed with the degrees of freedom
associated with nucleon field~($\psi$), two isoscalar meson fields
~($\sigma$ and $\omega_\mu$), isovector meson field~($\vec\rho_\mu$)
and photon field~($A_\mu$)~\cite{Ring1996,Vretenar2005,Meng2006a}.
Under ``mean-field" and ``no-sea" approximations, one can derive the
corresponding energy density functional, from which one finds
immediately the equation of motion for a single-nucleon orbital
$\psi_i(\bm{r})$ by variational principle,
\begin{equation}\label{DiracEq}
 \{\mathbf{\alpha}\cdot[\bm{p}- \bm{V} (\bm{r})]
  +\beta M^*(\bm{r})+V_0(\bm{r})\}\psi_i(\bm{r})
  =\epsilon_i\psi_i(\bm{r}),
\end{equation}
where $M^*(\bm{r})\equiv M+g_\sigma\sigma(\bm{r})$,
$M$ the mass of bare nucleon, $V_0(\bm{r})$ the time-like component of vector potential, and $\bm{V}(\bm{r})$ the space-like components of vector fields. The details about the solution of Dirac equation (\ref{DiracEq}) with time-odd fields can be found in Refs.~\cite{Yao2006,Li2009}.

In the configuration-fixed constrained calculation, the same configuration is guaranteed during the procedure of constraint calculation
with the help of ``parallel-transport"~\cite{Bengtsson1989}. In addition to the $\beta^2$-constrained calculation~\cite{Meng2006}, the constraints on the axial and triaxial mass quadrupole moments are also performed to obtained the potential energy surfaces (PES) in the two-dimensional $\beta$-$\gamma$ plane~\cite{Li2010}.


In the present calculations, each Dirac spinor $\psi_i(\bm{r})$ is expanded
in terms of a set of three-dimensional harmonic oscillator (HO) basis in Cartesian
coordinates with 12 major shells and the meson fields with 10 major shells. The pairing correlations are quenched by the unpaired valence nucleons in $^{105}$Rh and thus neglected here. The effective interaction parameter set PK1~\cite{Long2004} is applied. The center-of-mass (c.m.) correction~\cite{Long2004,Zhao2009}
is taken into account by
\begin{equation}
  \label{c.m.}
    E_{\mathrm{c.m.}}^{\mathrm{mic.}}=-\frac{1}{2MA}\langle
    \hat{\mathbf{P}}^2_{\mathrm{c.m.}}\rangle,
\end{equation}
where $\hat{\mathbf{P}}_{\mathrm{c.m.}}$ is the total momentum operator of a nucleus with mass
number $A$. In order to save time, the constrained triaxial RMF calculations without the time-odd fields are performed to search for the
triaxial deformation parameters and the valence nucleons
configuration favorable for chirality, which will be later confirmed by the calculations with time-odd fields.

 \begin{figure}[h!]
 \centering
 \includegraphics[width=8.5cm]{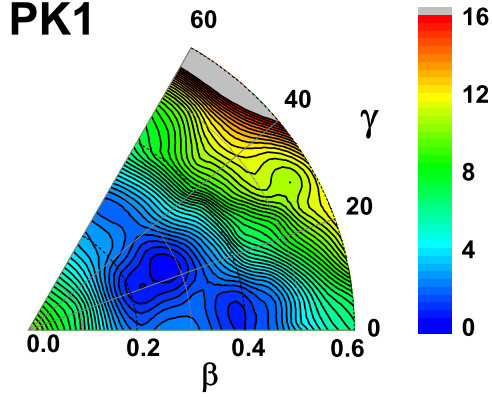}\vspace*{-0.5cm}
  \caption{(Color online) Contour plots of potential energy surface in $\beta$-$\gamma$ plane ($0\leqslant\gamma\leqslant60^\circ$) for $^{105}$Rh in
    constrained triaxial RMF calculations with effective interactions PK1~\cite{Long2004}. All energies are normalized with respect to the binding energy of the absolute minimum (in MeV). The energy separation between contour lines is 0.4 MeV.}
 \label{fig1}
 \end{figure}

The potential energy surface in the $\beta$-$\gamma$ plane
($0\leqslant\gamma\leqslant60^\circ$) for $^{105}$Rh in the
adiabatic constrained triaxial RMF calculations with PK1 is shown in
Fig.~\ref{fig1}. All energies are normalized with respect to the
binding energy of the absolute minimum, and the contours join points
on the surface with the same energy (in MeV). The energy separation
between contour lines is 0.4 MeV. From Fig.~\ref{fig1}, it shows
that the ground state of $^{105}$Rh is triaxially deformed with
$\beta=0.28$ and $\gamma=23^\circ$. The second minimum is located at
the area with $\beta\approx0.4$ and $\gamma\approx5^\circ$. The
energy difference between these two minima is about 0.5 MeV and
corresponding barrier height is about 1.5 MeV. The behavior of shape
coexistence is clearly shown.

\begin{figure}[h!]
 \centering
 \includegraphics[width=8.5cm]{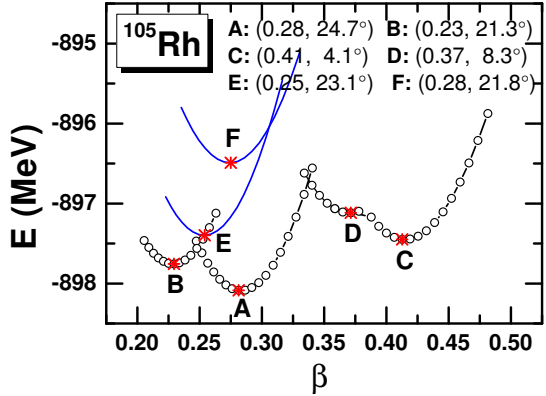}\vspace*{-0.5cm}
 \caption{(Color online) The energy surfaces in adiabatic (open circles) and
  configuration-fixed (solid lines) constrained triaxial RMF calculation using
  effective interaction PK1 for $^{105}$Rh. The minima in the energy surfaces
  for fixed configuration are represented as stars and labeled respectively
  as A, B, C, D, E, and F. Their corresponding triaxial deformation parameters
  $\beta$ and $\gamma$ are also given.}
 \label{fig2}
\end{figure}

The $\beta^2$-constrained calculations, in which the triaxial
deformation $\gamma$ is automatically obtained by minimizing the
energy, have also been performed. In Fig.~\ref{fig2}, the potential
energy surfaces as a function of $\beta$ in adiabatic (open circles)
constrained triaxial RMF calculation with PK1 for $^{105}$Rh are
given, where four minima observed in the potential energy surfaces
are labeled with A, B, C, and D respectively.

\begin{table*}
  \centering
  \tabcolsep=4pt
  \caption{The total energies $E_{\rm tot}$, triaxial deformation parameters
  $\beta,\gamma$ as well as their corresponding valence nucleon
  configurations of minima for A-F in the configuration-fixed constrained triaxial RMF calculations.}
  \begin{tabular}{ccccccc}
  \toprule
  \multirow{2}{*}{State} &\multicolumn{2}{c}{Configuration}   &\multirow{2}{*}{$E_{\rm tot}$}  &  \multirow{2}{*}{$\beta $} & \multirow{2}{*}{$\gamma $} \\ \cline{2-3}
        & Valence nucleons  & Unpaired nucleons & & & \\
  \hline
  A     &$\pi1g^{-3}_{9/2}\otimes\nu1h^{2}_{11/2}$ & $\pi1g^{-1}_{9/2}$ & -898.09                    & 0.28 &  24.7$^\circ$    \\
  B     &$\pi1g^{-3}_{9/2}\otimes\nu(1g^{-2}_{7/2}2d^4_{5/2})$ & $\pi1g^{-1}_{9/2}$ & -897.76         & 0.23  & 21.3$^\circ$     \\
  C     &$\pi(1g^{-4}_{9/2}2p^{-1}_{3/2}1g^{2}_{7/2})\otimes\nu(1g^{-2}_{7/2}3s^2_{1/2}1h^{2}_{11/2})$& $\pi2p^{-1}_{3/2}$ &-897.45   & 0.41  & 4.1$^\circ$     \\
  D     &$\pi(1g^{-4}_{9/2}1g^{1}_{7/2})\otimes\nu(1g^{-2}_{7/2}1h^{4}_{11/2})$ & $\pi1g^{1}_{7/2}$ & -897.12  & 0.37  & 8.3$^\circ$     \\ \hline
  E     &$\pi1g^{-3}_{9/2}\otimes\nu(1h^{1}_{11/2}2d^1_{5/2})$ & $\pi1g^{-1}_{9/2}\otimes\nu(1h^{1}_{11/2}2d^1_{5/2})$ & -897.40  & 0.25  & 23.1$^\circ$   \\
  F     &$\pi1g^{-3}_{9/2}\otimes\nu(1h^{1}_{11/2}1h^{1}_{11/2})$ &$\pi1g^{-1}_{9/2}\otimes\nu(1h^{1}_{11/2}1h^{1}_{11/2})$ & -896.40  & 0.28  & 21.8$^\circ$  \\
  \hline\hline
\end{tabular}\label{tab1}
\end{table*}

The total energies $E_{\rm tot}$, triaxial deformation parameters
$\beta,\gamma$ as well as their corresponding configurations of minima for A-D in the constrained triaxial RMF calculations are presented in Table~\ref{tab1}. Here, the state A represents the ground state, with triaxial deformation $\beta=0.28$, $\gamma=24.7^\circ$ and the corresponding valence nucleon configuration $\pi1g^{-3}_{9/2}\otimes\nu1h^{2}_{11/2}$. Note that the two of three proton holes in $1g_{9/2}$ orbital and two neutrons in the $1h_{11/2}$ orbital are pairwise, i.e., they occupy the degenerate time-reversal conjugate levels and don't contribute to the angular momentum. The corresponding unpaired nucleon configuration is $\pi1g^{-1}_{9/2}$. For state B, the triaxial deformation parameter is ($\beta=0.23$, $\gamma=21.3^\circ$), with corresponding valence nucleon configuration $\pi1g^{-3}_{9/2}\otimes\nu(1g^{-2}_{7/2}2d^4_{5/2})$ and unpaired nucleon configuration $\pi1g^{-1}_{9/2}$. Although A and B have different valence nucleon configurations, they have the same unpaired nucleon configuration.

The triaxial deformation parameters for excited state C and D are
($\beta=0.41$, $\gamma=8.3^\circ$) and ($\beta=0.37$,
$\gamma=8.3^\circ$), respectively. The valence nucleon configuration
for C is
$\pi(1g^{-4}_{9/2}2p^{-1}_{3/2}1g^{2}_{7/2})\otimes\nu(1g^{-2}_{7/2}3s^2_{1/2}1h^{2}_{11/2})$
and for D
$\pi(1g^{-4}_{9/2}1g^{1}_{7/2})\otimes\nu(1g^{-2}_{7/2}1h^{4}_{11/2})$,
while the unpaired nucleon configuration $\pi2p^{-1}_{3/2}$ for C
and $\pi1g^{1}_{7/2}$ for D. The quadrupole deformations of states C
and D are larger than states A and B, which is due to the
deformation driving orbital $\pi1g_{7/2}$ in states C and D. It
should be pointed out that states A, B, C and D don't have the
suitable particle-hole configurations for chirality.

In the Refs.~\cite{Timar2004,Alcantara2004}, the observed partners
bands with the configurations
$\pi1g^{-1}_{9/2}\otimes\nu1h^{1}_{11/2}(1g_{7/2},2d_{5/2})^1$ and
$\pi1g^{-1}_{9/2}\otimes\nu(1h^{1}_{11/2}1h^{1}_{11/2})$ are
respectively suggested as candidate chiral doublet bands. It is
interesting to examine the triaxial deformation of these states. For
this purpose, the configuration-fixed constrained calculation with
the suggested configurations are performed and their corresponding
energy surfaces are given in Fig.~\ref{fig2} in blue solid lines,
and the energies, configurations and triaxial deformation parameters
are listed in Table~\ref{tab1}. The excitation energy for minima E
is 0.69 MeV and for F 1.69 MeV. Furthermore, the triaxial
deformation suitable for chirality is found for E and F, namely
23.1$^\circ$ and 21.8$^\circ$ respectively, which together with the
corresponding high-$j$ proton hole and high-$j$ neutron particle
configurations will lead to the M$\chi$D phenomenon~\cite{Meng2006}.
It should be noted that the present RMF calculations are restricted to the non-rotating mean field only, while the TAC calculations with Woods--Saxon potentials have demonstrated that the rotating mean field will become chiral for configuration E at $\hbar \omega =0.20$ MeV (strong paring)~\cite{Alcantara2004} and for configuration F at $\hbar \omega =0.60$ MeV (zero pairing) or $0.45$ MeV (strong paring)~\cite{Timar2004}. Recently, the doublet bands with configuration F in $^{105}$Rh are also investigated by triaxial PRM calculations and the evolution of the chiral geometry with angular momentum is discussed~\cite{Qi2011}.

\begin{figure*}
\centering
\includegraphics[width=8.5cm]{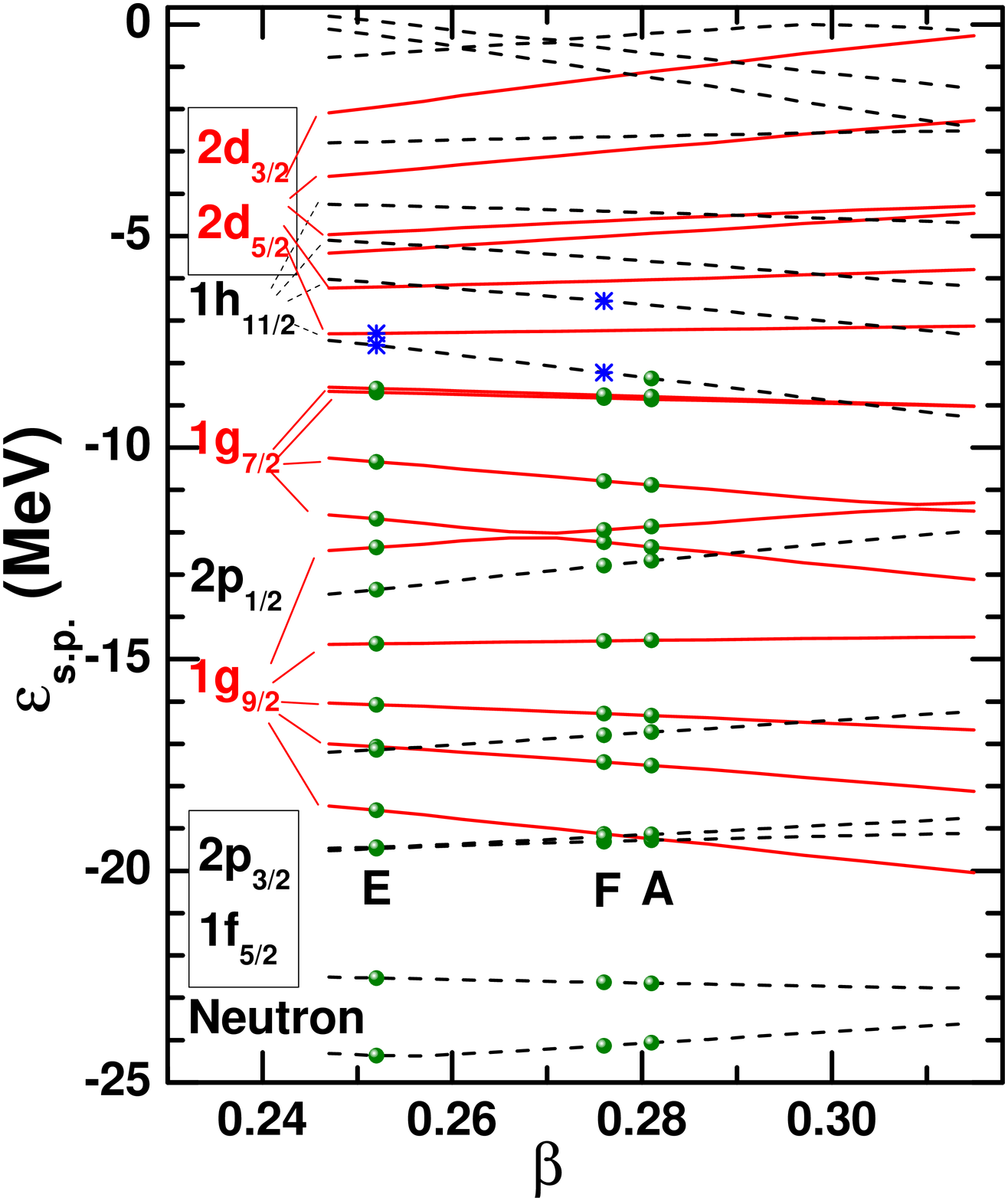}
\includegraphics[width=8.5cm]{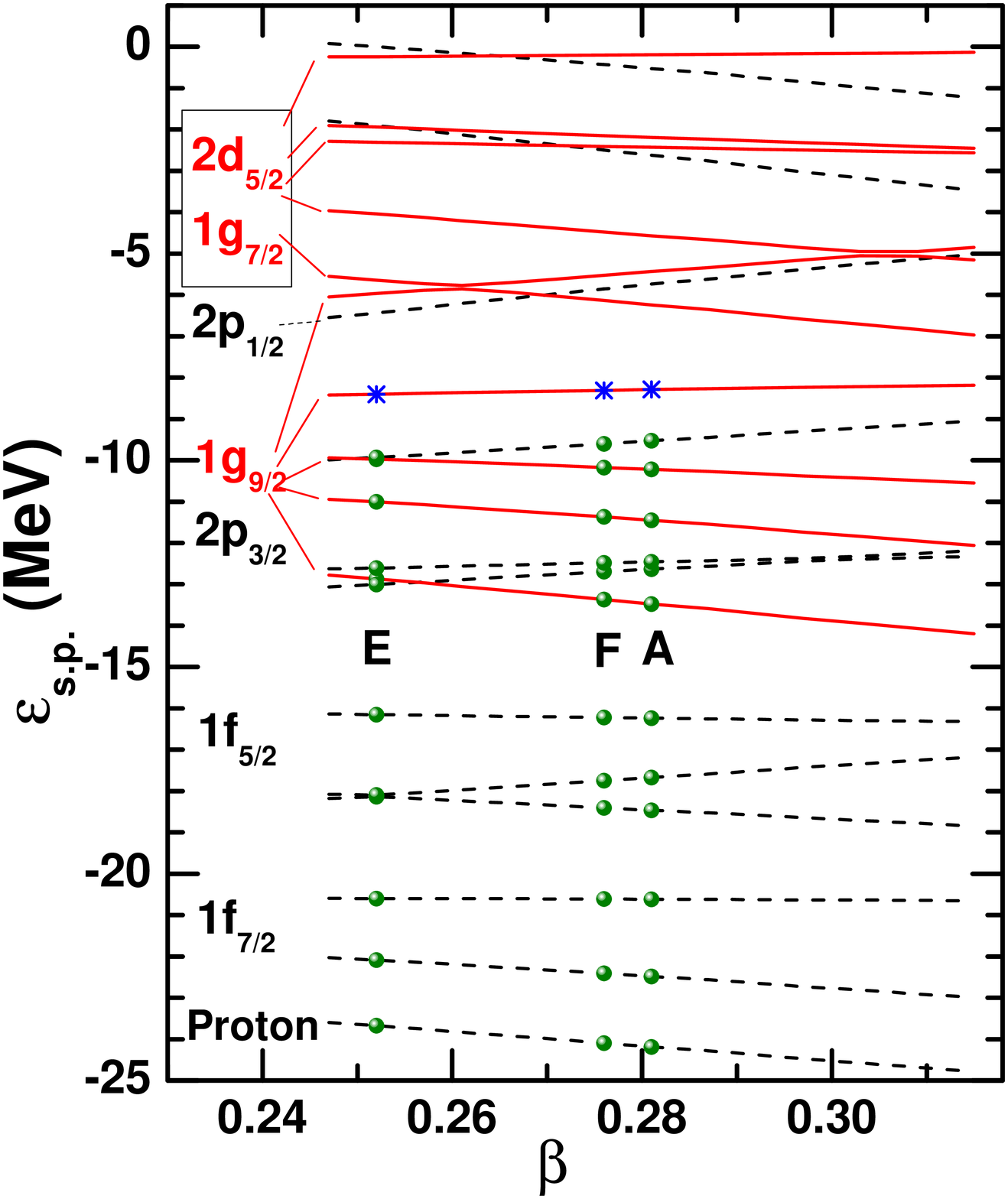}
\caption{(Color online) Neutron and proton single-particle levels obtained in constrained triaxial RMF calculations with
PK1 as functions of deformation $\beta$. Positive (negative) parity
states are marked by solid (dashed) lines. Occupations corresponding
to the minima in Fig.~\ref{fig2} are represented by filled circles
(two particles) and stars (one particle).}
\label{fig3}
\end{figure*}

The neutron and proton single-particle levels as functions of
deformation $\beta$ are given in Fig.~\ref{fig3}, obtained in the
adiabatic constrained triaxial RMF calculations for the region
$0.247<\beta<0.315$. The occupations of neutron and proton for A, E
and F are given in Fig.~\ref{fig3}(a) and Fig.~\ref{fig3}(b)
respectively. The single-particle levels with positive (negative)
parity are marked by solid (dashed) lines, and occupations are
represented by filled circles (two particles) and stars (one
particle). It should be pointed out that every single-particle level
is degenerated with opposite simplex quantum number~\cite{Peng2008}.
It is easy to see that for ground state A, one valence proton
occupies the $1g_{9/2}$ orbit and the last two neutrons occupy the
degenerated time reversal conjugate orbit $1h_{11/2}$. For the positive parity state F, the two $1h_{11/2} $ neutrons are unpaired (in contrast to the state A), i.e., they occupy two different $1h_{11/2} $ levels. For the negative parity state E, the two unpaired neutrons occupy the $1h_{11/2} $ and $2d_{5/2} $ levels. It is
interesting to note that states E and F compete with each other in
energy. However, due to different parities, states E and F do not
mix up and could contribute the
M$\chi$D phenomenon~\cite{Meng2006}.

Similar as in Ref.~\cite{Yao2009}, the calculations with time-odd
fields are also performed to confirm the above discussions. In
Table~\ref{tab2}, the calculated triaxial deformation parameters
$\beta,\gamma$, total energies $E_{\mathrm{tot}}$, and the
excitation energies $E_\mathrm{x}$ for states E and F with and
without time-odd fields are given and compared with the experimental
bandhead energies of two rotational bands. The effects of the
time-odd fields on the triaxial deformation parameters
$\beta,\gamma$ are negligible. Their contribution for the total
energy are respectively $-0.05$ MeV for state A,  $-0.3$ MeV for
state E and $-0.36$ MeV for state F.

The experimental spin, parity and excitation energies of bandheads
for two rotational bands based on the configurations of E and F are
${\frac{15}{2}}^{-}$, $2.417$ MeV and ${\frac{23}{2}}^{+}$, $2.982$
MeV respectively. It can be seen that the displacement for the
energies of two bandheads (0.57 MeV) has been reasonable reproduced
by the RMF calculations (1.0 and 1.06 MeV).

\begin{table*}
  \centering
  \tabcolsep=8pt
  \caption{The triaxial deformation parameters
  $\beta,\gamma$, total energies $E_{\mathrm{tot}}$ and the excitation energies $E_\mathrm{x}$ for states E and F calculated with and without time-odd fields, compared with the experimental bandhead energies of two rotational bands based on the configurations of E and F. The spin and parity $I^\pi$ for ground state and band head of two rotational bands are also given.}
  \begin{tabular}{cccccccccc}
  \toprule
  \multirow{2}*{State} & \multicolumn{2}{c}{Time-even} && \multicolumn{2}{c}{Time-odd} && \multicolumn{2}{c}{exp} \\ \cline{2-3} \cline{5-6} \cline{8-9}
    & $(\beta,\gamma)$ & $E_\mathrm{x}$($E_{\mathrm{tot}}$)  && $(\beta,\gamma)$ & $E_\mathrm{x}$($E_{\mathrm{tot}}$)  & & $I^\pi$ & $E_\mathrm{x}$  \\
  \hline
  A & (0.28,24.7$^\circ$) & ~~~~~(-898.09) & & (0.28,24.9$^\circ$) & ~~~~~(-898.14) & &                      ${\frac{7}{2}}^{+}$&      \\
  E & (0.25,23.1$^\circ$) & 0.69(-897.40) & & (0.25,23.3$^\circ$) & 0.37(-897.76) & & ${\frac{15}{2}}^{-}$ & 2.417\\
  F & (0.28,21.8$^\circ$) & 1.69(-896.40) & & (0.28,22.1$^\circ$) & 1.43(-896.70) & & ${\frac{23}{2}}^{+}$ & 2.982\\
  \hline\hline
\end{tabular}\label{tab2}
\end{table*}

It is necessary to note here that in $^{105}$Rh, the suggested configuration $\pi1g^{-1}_{9/2}\otimes\nu1h^{1}_{11/2}(1g_{7/2},2d_{5/2})^1$ for the candidate chiral doublet bands (band 7 and 8 in Ref.~\cite{Alcantara2004}) involves the orbits of pseudospin doublet states $(1g_{7/2}, 2d_{5/2})$. So a competing interpretation of band 7 and 8 includes the pseudospin doublet bands~\cite{Meng2010}. Further efforts are needed to address this point.

In summary, adiabatic and configuration-fixed constrained triaxial
RMF approaches have been applied to investigate the M$\chi$D
candidate nucleus $^{105}$Rh. According to the suggested high-$j$
proton hole and high-$j$ neutron particle configurations, their
triaxial deformations favorable for the construction of the chiral doublet
bands are obtained from the configuration-fixed constrained triaxial
RMF calculations. The existence of M$\chi$D phenomenon is
expected in $^{105}$Rh. Future efforts should be made to investigate whether the rotating mean field will attain chirality or not.

\begin{acknowledgments}
This work is partly supported by Major
State Basic Research Developing Program 2007CB815000 and the National
Natural Science Foundation of China under Grant Nos. 10975007, 10975008 and 110050691.
\end{acknowledgments}



\end{document}